\documentclass[preprint,superscriptaddress,showkeys,showpacs,
tightenlines,fleqn,nofootinbib,prd,byrevtex]{revtex4} 
\usepackage{bm}
\usepackage{graphicx}
\usepackage{epstopdf}
\begin{document}      
\preprint{YITP-08-26}
\preprint{INHA-NTG-11/2008}
\title{Pion weak decay constant at finite density \\ 
from the instanton vacuum}       
%-------------------------------------------------
\author{Seung-il Nam}
\email[E-mail: ]{sinam@yukawa.kyoto-u.ac.jp}
\affiliation{Yukawa Institute for Theoretical Physics (YITP), Kyoto
University, Kyoto 606-8502, Japan} 
%-------------------------------------------------
\author{Hyun-Chul Kim}
\email[E-mail: ]{hchkim@inha.ac.kr}
\affiliation{Department of Physics, Inha University, Incheon 402-751,
  Korea.}  
%-------------------------------------------------
\date{May 2008}
\begin{abstract}  
We investigate the pion weak decay constant ($F_\pi$) and pion mass 
($m_\pi$) at finite density within the framework of the nonlocal
chiral quark model from the instanton vacuum with the finite
quark-number chemical potential ($\mu$) taken into account.  We mainly
focus on the Nambu-Goldstone phase below the critical value of the
chemical potential $\mu_c\approx320$ MeV, which is determined
consistently within the present framework.  The breakdown of Lorentz
invariance at finite density being considered, the time 
($F^t_\pi$) and space ($F^s_\pi$) components are computed separately,
and the corresponding results turn out to be: $F^t_\pi=82.96$ MeV and
$F^s_\pi=80.29$ MeV at $\mu_c$, respectively.  Using the in-medium 
Gell-Mann-Oakes-Renner (GOR) relation, we show that the pion mass
increases by about $15\%$ at $\mu_c$.   
\end{abstract} 
\maketitle
%--------------------------------------------------
\section{Introduction}
%--------------------------------------------------
The in-medium modifications of the pion have been one of the most
interesting issues both in experimental and theoretical hadron
physics.  The pion is identified as the Goldstone boson arising
from the spontaneous breakdown of chiral symmetry (SB$\chi$S) which is
essential in describing low-energy hadronic phenomena.  Since chiral
symmetry is expected to be restored at high temperature and
density, the changes of pion properties in medium will provide crucial 
information on the restoration of chiral symmetry.  Among the
properties of the pion, its weak decay constant ($F_\pi$) and mass
($m_\pi$) are the most important quantities, since they are deeply 
related to the SB$\chi$S: The $F_\pi$ and $m_\pi$ are related to the
quark mass and chiral condensate via the Gell-Mann-Oakes-Renner (GOR)  
relation~\cite{GellMann:1968rz}, so that the in-medium modification of
the $F_\pi$ and $m_\pi$ will give a key clue for understanding the
mechanism of the chiral symmetry restoration in matter. 

Experimentally, the modifications of the $F_\pi$ and $m_\pi$ can be 
measured from deeply bound pionic atoms, the $s$-wave pion-nucleus
interaction being considered in medium~\cite{Suzuki:2002ae,
Gilg:1999qa,Itahashi:1999qb,Geissel:2002ur,Yamazaki:1996zb}, 
as suggested by Refs.~\cite{Toki:1989wq,Toki:1990fc,Hirenzaki:1991us}
and other related works~\cite{Kolomeitsev:2002gc,Toki:2005cc}.
The in-medium change of the pion mass can be also probed in the
Drell-Yan process~\cite{Dieperink:1997iv,Brown:1993sua}.  

There has been a great deal of theoretical work on the in-medium
modifications of the $F_\pi$ and $m_\pi$.  For example, meson-baryon 
chiral perturbation theory ($\chi$PT) and models with chiral 
symmetry were applied for this purpose~\cite{Kirchbach:1993ep,Bernard:1987ir,Hatsuda:1994pi,Thorsson:1995rj,Wirzba:1995sh,Mallik:2003dt,Kirchbach:1997rk,Hatsuda:1994pi,Meissner:2001gz,Kim:2003tp,Kaiser:2001bx,Delorme:1996cc,Park:2001ht,
Kaiser:2007nv}. Since the Lorentz invariance is broken in 
medium, one has to study the space and time components of the pion
weak decay constant separately.  In in-medium
$\chi$PT~\cite{Kirchbach:1997rk,Meissner:2001gz} for instance,   
the magnitude of its space component $F^s_\pi$ was shown to be about 
four times smaller than that of the time component $F^t_\pi$ at   
normal nuclear density $\rho_0\approx0.17\,\mathrm{fm}^{-3}$.
Moreover, the chiral condensate varies sizably in
medium~\cite{Hatsuda:1994pi,Meissner:2001gz,Bernard:1987ir,Thorsson:1995rj,Wirzba:1995sh,Kaiser:2007nv}.  It was also shown that within these
approaches, the contribution of the intermediate $\Delta$ isobar  
explains the suppression of the $F^s_\pi$ in comparison with the
$F_\pi^t$.  In the QCD sum rules, it was discussed that dimension-five   
operators are responsible for making splitting between $F^t_\pi$ and
$F^s_\pi$, and the contribution of the intermediate $\Delta$ isobar
contribution makes $F^s_\pi$ much smaller than $F^t_\pi$, while they
are approximately in the same order without it~\cite{Kim:2003tp}.    
Taking into account the experiments conducted in
Ref.~\cite{Yamazaki:1996zb,Yamazaki:1997na} and using effective
potential models, Refs.~\cite{Waas:1997kg,Friedman:1998ed} have
studied the deeply bound pionic atoms and have shown that the encoded
pion-mass modification turned out to be also small (about $10\%$).  

In the present work, we would like to investigate the modifications of 
the $F_\pi$ and $m_\pi$ at finite quark-number chemical potential
($\mu\ne0$) but at zero temperature ($T=0$), employing the nonlocal
chiral quark model (NL$\chi$QM) that is derived from the nontrivial
instanton vacuum~\cite{Diakonov:1985eg} with the finite quark-number
chemical potential considered in the $N_c$ limit~\cite{Carter:1998ji}.
The NL$\chi$QM from the instanton vacuum is characterized by the
average instanton size $\bar{\rho} \approx1/3$ fm and inter-instanton
distance $\bar{R} \approx 1$ fm.  The scale of the model is given by
the average instanton size, {\it i.e.}
$\Lambda\approx1/\bar{\rho}\approx 600$ MeV.  We already have
successfully applied this modified NL$\chi$QM to the pion
electromagnetic form factor~\cite{Nam:2008fe} and magnetic
susceptibility of the QCD vacuum~\cite{Nam:2008ff}.  As done
previously, we will mainly focus on the Nambu-Goldstone (NG) phase
below $\mu=\mu_c\approx320$ MeV, which is close to $\rho_0$.  
We will show in the present work that the time and space components of
the pion weak decay constant will turn out to be $F^t_\pi=82.96$ MeV
and $F^s_\pi=80.29$ MeV at $\mu_c$, which are about $13\sim16\%$
smaller than that in free space ($F_\pi=93$ MeV).  The results are
compatible with those obtained in other models, though the result for
the $F_\pi^s$ seems to be larger than those from $\chi$PT and from the
QCD sum rules.  Using the GOR relation that is satisfied within the 
model~\cite{Diakonov:1985eg}, we estimate the pion mass shift,
resulting in about $15\%$ increase at $\mu_c$.    

The present work is organized as follows: In Section II, we briefly 
review the general formalism in the NL$\chi$QM at finite density.  In
Section III, we discuss the phase structures in the present framework.
In Section IV, the numerical results are given and   
discussed.  The final Section is devoted to summary and conclusions.

%-------------------------------------------------
\section{Nonlocal chiral quark model at finite density}
%-------------------------------------------------
The Dirac equation in the presence of the finite quark-number chemical
potential ($\mu$) in the instanton (anti-instanton) background field
can be written as follows:   
%EQAUTION>>>
\begin{equation}
\label{eq:MDE}
\left[i\rlap{/}{\partial}-i\gamma_4\mu
-\rlap{/}{A}_{I\bar{I}}\right]\Psi^{(n)}_{I\bar{I}}
=\lambda_n\Psi^{(n)}_{I\bar{I}}.
\end{equation}
%EQUATION<<<
In the present work, we work in Euclidean space and assume the chiral limit
($m_q=m_\mathrm{u}=m_\mathrm{d}=0$).  The subscript $(\bar{I})I$ stands 
for the (anti)instanton contribution, and we use a singular-gauge
instanton solution: 
%EQUATION>>>
\begin{eqnarray}
\label{eq:instanton}
A^{\alpha}_{I\bar{I}\mu}(x)
=\frac{2\bar{\eta}^{\alpha\nu}_{\mu}\bar{\rho}^2x_{\nu}}
{x^2(x^2+\bar{\rho}^2)},
\end{eqnarray}
%EQUATION<<<
where $\eta^{\alpha\nu}_{\mu}$ and $\bar{\rho}$ denote the 't Hooft
symbol and average instanton size, respectively. The quark zero-mode 
solution can be obtained in the presence of the $\mu$ as follows: 
%EQAUTION>>>
\begin{equation}
\label{eq:ZM}
\left[i\rlap{/}{\partial}-i\gamma_4\mu
-\rlap{/}{A}_{I\bar{I}}\right]\Psi^{(0)}_{I\bar{I}}
=0.
\end{equation}
%EQUATION<<<
The explicit form of $\Psi^{(0)}$ can be found in
Ref.~\cite{Carter:1998ji}.  The effective chiral action can be
constructed by the would-be zero mode of Eq.(\ref{eq:ZM}).  The quark
propagator in one instanton background goes to infinity in the chiral
limit ($m_q\to 0$)~\footnote{Note that the chiral limit will be taken
  after the low-energy effective partition function is obtained.}:
%EQUATION>>>
\begin{equation}
\label{eq:prop1}
S_{I\bar{I}}(x,y)=\langle \psi(x)\psi(y)^\dagger\rangle = -\sum_n 
\frac{\Psi_{I\bar{I}}^{(n)}(x)\Psi_{I\bar{I}}^{(n)\dagger} 
}{\lambda_n+im_q} = S_{I\bar{I}}'(x,y) -
\frac{\Psi^{(0)}_{I\bar{I}}(x) \Psi^{(0)\dagger}_{I\bar{I}}(y)}{im_q}, 
\end{equation}
where $S_{I\bar{I}}'$ denotes the non-zero mode contribution.
Since the zero-mode contribution dominates in Eq.(\ref{eq:prop1}) at
small momenta ($p\lesssim 1/\bar{\rho}$) whereas it is reduced to the 
free quark propagator at large momenta ($p\gg 1/\bar{\rho}$),       
the quark propagator can be approximated as  
\begin{equation}
\label{eq:prop11}
S_{I\bar{I}}(x,y) \approx S_0 -
\frac{\Psi^{(0)}_{I\bar{I}} \Psi^{(0)\dagger}_{I\bar{I}}}{im_q},
\end{equation}
%EQAUTION<<<
where $S_0$ is a free quark propagator defined as
$(i\rlap{/}{\partial} - i\gamma_4\mu)^{-1}$.  Starting from the
zero-mode approximation in Eq.(\ref{eq:prop11}), one can derive the
quark propagator in the instanton
ensemble~\cite{Diakonov:1985eg,Carter:1998ji}: 
%EQUATION>>>
\begin{equation}
\label{eq:prop2}
S=\frac{1}{i\rlap{/}{\partial}-i\gamma_4\mu+iM(i\partial,\mu)},
\end{equation}
%EQUATION<<<
where $M$ denotes the momentum-dependent and $\mu$-dependent quark
mass that arises from the Fourier transform of the quark zero-mode
solution:  
%EQUATION>>>
\begin{equation}
\label{eq:MDQM}
M(\bar{k})=M_0(\mu)\bar{k}^2\psi^2(\bar{k}).  
\end{equation}
%EQUATION<<<
Here, $\bar{k}=(\vec{k},k_4+i\mu)$.  The $M_0$ is the constituent
quark mass at $k^2=0$, which depends on $\mu$.  It will be determined 
consistently within the model.  The analytical expressions for
$\psi_4$ and $\vec{\psi}$ are given in Appendix~\cite{Carter:1998ji}.   

The low-energy effective partition function of the NL$\chi$QM with
$\mu\ne0$ can be written as follows: 
%EQAUTION>>>
\begin{equation}
\label{eq:PF}
\mathcal{Z}_{\rm eff}=\int{d\lambda}\,{D\psi}\,{D\psi^{\dagger}}
\exp\left[\int d^4x\psi^{\dagger}(i\rlap{/}{\partial}-i\gamma_4\mu)\psi
+\lambda(Y^++Y^-)+N\left(\ln\frac{N}{\lambda
    V\mathcal{M}}-1\right)\right], 
\end{equation}
%EQAUTION<<<
where $V$ indicates the four-dimensional volume, whereas $N$ 
represents the average number of the (anti)instantons,
$N=(N_++N_-)/2$.  The variational parameter $\lambda$ plays a role of
a Lagrangian multiplier. The $Y^{\pm}$ stands for the $2N_f$-'t Hooft
interaction in the instanton background with nonzero $\mu$. The
parameter $\mathcal{M}$ is required to make the argument of the
logarithm dimensionless. All calculations are performed to order 
$\mathcal{O}(\lambda)$.  
%-------------------------------------------------
\section{Phase structures: NG and CSC phases}
\begin{figure}[t]
\begin{tabular}{cc}
\includegraphics[width=7.5cm]{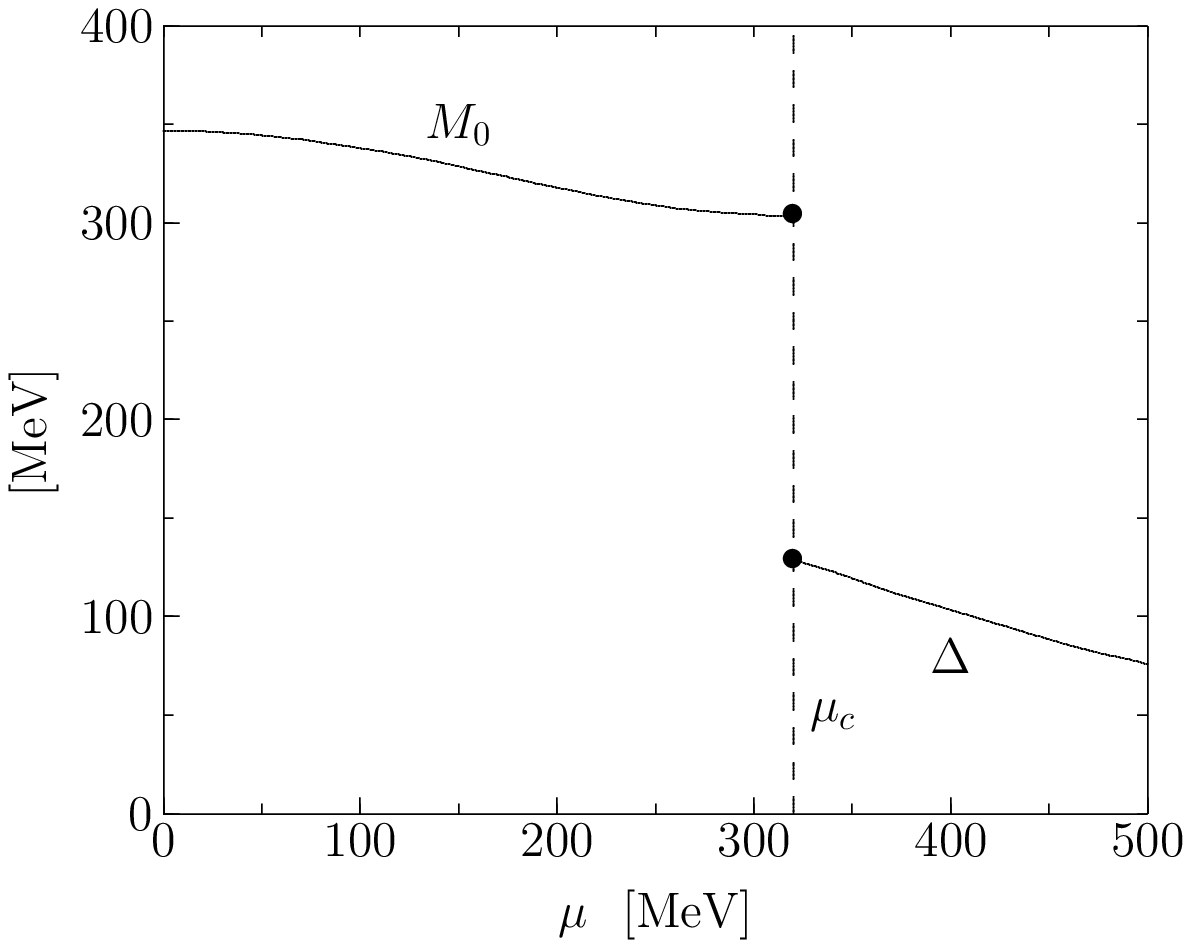}
\includegraphics[width=7.5cm]{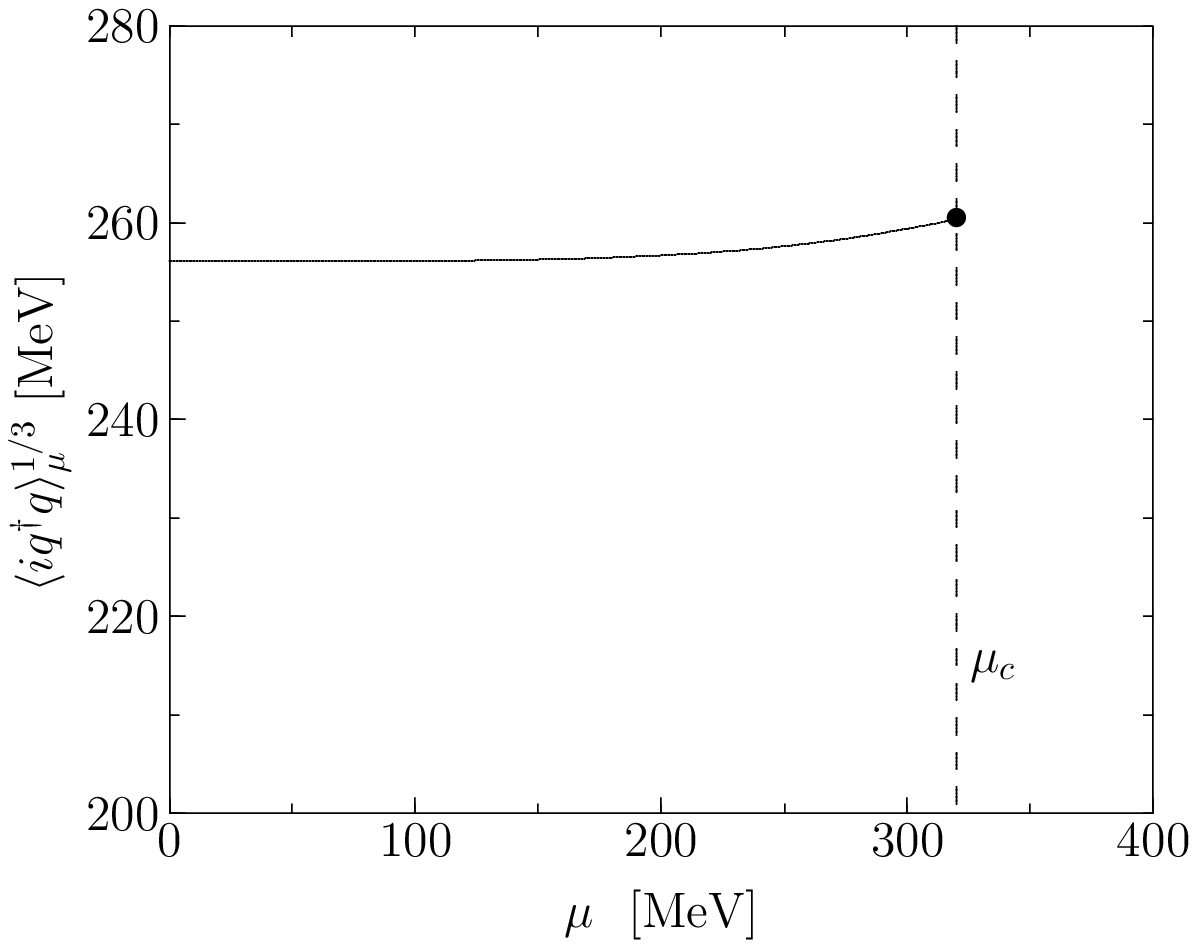}
\end{tabular}
\caption{$M_0$ and $\Delta$ (left), and
  $\langle{iq^{\dagger}q}\rangle^{\frac{1}{3}}$ (right) as
  functions of the quark-number chemical potential $\mu$ up to
  $\mu=\mu_c\approx320$ MeV. Vertical dashed lines indicate the
  critical density, 
  $\mu_c\approx320$ MeV.}       
\label{fig0}
\end{figure}

%-------------------------------------------------
In this Section, we want to discuss the phase structure for $\mu\ne0$
and $T=0$. Since we are interested in the case of $N_f=2$ and $N_c=3$,
there are various phase structures characterized by the different
order parameters $g$ and $f$ for the NG and CSC phases,
respectively~\cite{Carter:1998ji}. They can be computed from the quark
loops of the normal ($G$) and abnormal ($F$) quark propagators, which
correspond to the Dyson-Schwinger-Gorkov (DSG) equations: 
%EQUATION>>>
\begin{eqnarray}
\label{eq:DSGE}
Z(k)&=&1-G(k)A(k,\mu)M_0,
\cr
G(k)&=&Z(k)\psi^2(p)M_0,
\cr
F(k)&=&2Z(-k)\psi_{\mu}(k,\mu)\psi^{\mu}(-k,\mu)\Delta,
\end{eqnarray}
%EQUATION<<<
where the vertex functions are defined by
%EQUATION>>>
\begin{eqnarray}
\label{eq:VF}
A(k,\mu)&=&(k+i\mu)^2\psi^2(k,\mu),
\cr
B(k,\mu)&=&(k^2+\mu^2)\psi_{\mu}(k,\mu)\psi^{\mu}(-k,\mu)
+(k+i\mu)_{\mu}\psi^{\mu}(k,\mu)
(k-i\mu)_{\nu}\psi^{\nu}(-k,\mu)
\cr
&-&(k+i\mu)_{\mu}\psi^{\mu}(-k,\mu)
(k-i\mu)_{\nu}\psi^{\nu}(k,\mu).
\end{eqnarray}
%EQUATION<<<
Here, $\Delta$ stands for the diquark energy gap, corresponding to the
diquark correlation.  Note that we consider only the pure NG and CSC
phases here for simplicity.  Hence, the metastable mixed phases of the
NG and CSC are not taken into account.  The two phases are then
characterized by $g\ne0$ and $f=0$ for the NG phase and vice versa for 
the CSC one.  Using Eqs.~(\ref{eq:DSGE}) and (\ref{eq:VF}), the
condensates $f$ and $g$ can be written as follows:  
%EQUATION>>>
\begin{equation}
\label{eq:fg}
g(\mu)=\frac{\lambda{M_0}}{N^2_c-1}
\int\frac{d^4k}{(2\pi)^4}
\frac{\alpha(k,\mu)}{1+\alpha(k,\mu)M^2_0},
\,\,\,\,
f(\mu)=\frac{2\lambda\Delta}{N^2_c-1}
\int\frac{d^4k}{(2\pi)^4}
\frac{\beta(k,\mu)}{1+4\beta(k,\mu)\Delta^2},
\end{equation}
%EQUATION<<< 
where 
%EQUATION>>>
\begin{equation}
\label{eq:albe}
\alpha(k,\mu)=A(k,\mu)\psi^2(k,\mu),\,\,\,\,
\beta(k,\mu)=B(k,\mu)\psi_{\mu}(k,\mu)\psi^{\mu}(-k,\mu).
\end{equation}
%EQUATION<<<
In turn, $M_0$ and $\Delta$ can be also expressed in terms of $g$ and
$f$: 
%EQUATION>>>
\begin{equation}
\label{eq:fffff}
M_0=\left(2N_c-\frac{2}{N_c}\right)g(\mu),\,\,\,\,
\Delta=\left(1+\frac{1}{N_c} \right)f(\mu).
\end{equation}
%EQUATION<<<
Differentiating the partition function of Eq.~(\ref{eq:PF}) with
respect to $\lambda$, we can obtain the following saddle-point
equation: 
%EQUATION>>>
\begin{equation}
\label{eq:SPE}
\frac{N}{V} = \lambda\langle{Y^++Y^-}\rangle =
\frac{4(N^2_c-1)}{\lambda} \int\frac{d^4k}{(2\pi)^4}
\left[N_cM_0G(k)+4\Delta F(k)\right].
\end{equation}
%EQUATION<<<
Since $\langle{Y^++Y^-}\rangle$ corresponds to a $\infty$-shape
quark-loop integral with $G$ and $F$ for $N_f=2$, it can be rewritten
in terms of $f$ and $g$ as given above. Note that there is one caveat:
We assume that there is no density dependence in the instanton packing
fraction $N/V$ for the $\mathcal{O}(\lambda)$. Thus, we use $N/V
\approx(200\, \mathrm{MeV})^4$ for $\mu\ge0$.  Inserting
Eq.~(\ref{eq:fg}) into Eq.~(\ref{eq:SPE}) for the NG and CSC phase  
regions, we can obtain $M_0$ and $\Delta$ numerically from the
saddle-point equation.  In the left panel of Fig.~\ref{fig0}, we draw 
$M_0$ and $\Delta$ as a function of $\mu$.  The critical
density is determined by the following condition~\cite{Carter:1998ji}: 
%EQUATION>>>
\begin{equation}
\label{eq:fog}
\frac{f(\mu)}{g(\mu)}\Bigg|_{\mu=\mu_c}
=\left[\frac{N_c(N_c-1)}{2}\right]^{\frac{1}{2}},
\end{equation}
%EQUATION<<<
where $\mu_c$ becomes about $320$ MeV for $\bar{R}\approx1$ fm and
$\bar{\rho}=\frac{1}{3}$ fm.  In Fig~\ref{fig0}, $\mu_c$ is indicated by
vertical dashed lines.  Note that $M_0$ decreases as $\mu$ increases
and disappears for the region beyond $\mu_c$.  The diquark energy gap
starts to exist ($\Delta\approx120$ MeV) above the critical
point.  These numerical results are basically obtained in
Ref.~\cite{Carter:1998ji}.  

%FIGURE>>>
%FIGURE<<<
Similarly, we can also derive the chiral condensate from the partition
function, which is another order parameter for the NG phase: 
%EQUATION>>>
\begin{equation}
\label{eq:CC}
\langle iq^{\dagger}q\rangle_{\mu}=4N_c\int
\frac{d^4k}{(2\pi)^4}\left[\frac{M(k,\mu)}
{(k+i\mu)^2+M^2(k,\mu)}\right].
\end{equation}
%EQUATION<<<
In the right panel of Fig.~\ref{fig0}, we show the numerical results 
for the chiral condensate with respect to the $\mu$.  It turns out
that it increase slowly as $\mu$ does.  At the critical point, it
becomes larger than its vacuum value by about $4\sim5\%$.  Thus, we 
can conclude that the chiral condensate remains nearly the same as 
that for free space within the NG phase, as pointed out in
Refs.~\cite{Buballa:2003qv,Cabibbo:1975ig}.  A similar tendency was
also shown in the Nambu-Jona-Lasinio model (NJL)~\cite{Chang:2005iv, 
Miyamura:2002en} and in the two-color lattice
simulation~\cite{Nishida:2003tg}.  However, note that the present
result is rather different from those computed in in-medium 
$\chi$PT and other effective model calcultaions~\cite{Hatsuda:1994pi,Meissner:2001gz,Bernard:1987ir,Thorsson:1995rj,Wirzba:1995sh,Kaiser:2007nv} in which it was shown that the condensate decreases as nuclear density increases. 

%-------------------------------------------------
\section{Pion weak decay constant in medium}
%-------------------------------------------------
The pion weak decay constant $F_\pi$ can be defined as the following 
transition matrix element: 
%EQUATION>>>
\begin{equation}
\label{eq:PCAC}
\langle0|A^a_{\mu}(x)|\pi^b(P)\rangle=i\sqrt{2}F_\pi\delta^{ab}P_{\mu}
e^{-iP\cdot x},
\end{equation}
%EQUATION<<<
where $A^a_{\mu}$ and $P_{\mu}$ denote the axial-vector current, 
$\psi^{\dagger}\gamma_5\gamma_{\mu}\frac{\tau^a}{2}\psi$, and the pion
on-shell momentum $P^2=m^2_\pi$, respectively.  Since the Lorentz
invariance of the matrix element is broken at finite density,
Eq.~(\ref{eq:PCAC}) should be decomposed into the space and time parts
as follows: 
%EQUATION>>>
\begin{equation}
\label{eq:PCACst}
\langle0|A_i^a(x)|\pi^b(P)\rangle=i\sqrt{2}F^s_\pi
\delta^{ab} P_i e^{-iP\cdot x},\,\,\,\,
\langle0|A^a_4(x)|\pi^b(P)\rangle=i\sqrt{2}F^t_\pi\delta^{ab}
P_4e^{-iP\cdot x}.
\end{equation}
%EQUATION<<<

From the low-energy effective partition function given in
Eq.~(\ref{eq:PF}), one can derive an effective chiral action in terms
of the quarks and NG boson fields $\pi^a$ with the bosonization
carried out~\cite{Diakonov:2002fq}: 
%EQUATION>>>
\begin{equation}
\label{eq:ECA0}
\mathcal{S}_{\mathrm{eff}} [\pi,\mu] = -\mathrm{Sp}\ln
\left[i\rlap{/}{\bar{\partial}} + i\sqrt{M(i\bar{\partial})}
U_5\sqrt{M(i\bar{\partial})}\right],
\end{equation}
%EQUATION<<<
where $\mathrm{Sp}$ indicates the functional trace over color ($c$),
flavor ($f$) and Dirac ($\gamma$) spaces, {\it i.e.} $\int
d^4x\,\mathrm{Tr}_{c,f,\gamma}\langle\cdots\rangle$.  The
density-modified covariant derivative is defined as
$i\bar{\partial}_{\mu}=i\partial_{\mu}-i(\vec{0},\mu)$.  The $U_5$
stands for the nonlinear NG boson field: 
%EQUATION>>>
\begin{equation}
\label{eq:U5}
U_5=\frac{1+\gamma_5}{2}\,U+\frac{1-\gamma_5}{2}\,U^{\dagger}
=\exp\left(\frac{i\gamma_5\bm\pi\cdot\bm\tau}{F_0}\right).
\end{equation}
%EQUATION<<<
Here $F_0$ and $\tau^a$ denote a generic normalization constant for
$\pi$ and the Pauli matrix, respectively.  We use the following
expression for $\pi^a$:   
%EQUATION>>>
\begin{equation}
\label{eq:fr}
{\bm \pi}\cdot\bm\tau=\left(
\begin{array}{cc}
\pi^0&\sqrt{2}\pi^+\\
\sqrt{2}\pi^-&\pi^0\\
\end{array}
\right).
\end{equation}
%EQUATION<<<
We would like to emphasize that the pion field $\pi^a$ in
Eq.~(\ref{eq:ECA0}) is not a physical one, since it is introduced as
an auxiliary field in the bosonization.  Considering the field
renormalization, we can write $\pi^a_\mathrm{phy}$ as
%EQUATION>>>
\begin{equation}
\label{eq:pp}
\pi^a_\mathrm{phy}=\frac{1}{C_r}\pi^a,
\end{equation}
%EQUATION<<<
where $C_r$ stands for a certain field-renormalization
parameter. In the presence of the quark-number chemical potential, the 
physical NG-boson field itself can be modified as the Migdal field in
meson-nucleon $\chi$PT with nonzero baryon density
$\rho$~\cite{Kirchbach:1996xy}.  Thus, $C_r$ is given in principle as
a function of $\mu$, {\it i.e.} $C_r(\mu)$, which depends on an appropriate
renormalization scale.  We, however, have assumed that the generic
normalization constant $F_0$ does not change at finite density. 

In order to compute the pion weak decay constant in
Eq.~(\ref{eq:PCAC}), we rewrite the effective chiral action in the 
presence of an external axial-vector source $J^a_{5\mu}$: 
%EQUATION>>>
\begin{eqnarray}
\label{eq:ECA}
\mathcal{S}_\mathrm{eff}[\pi,\mu,J^a_{5\mu}]&=&
-\mathrm{Sp}\ln
\left[i\rlap{/}{\bar{\partial}}
+\gamma_5\gamma^{\mu}\frac{\tau^a}{2}J^a_{5\mu}
+\sqrt{M(i\bar{\partial},J^a_{5\mu})} U_5
\sqrt{M(i\bar{\partial},J^a_{5\mu})} \right]. 
\end{eqnarray}
%EQUATION<<<
Since it is well known that the gauge invariance is broken in the
presence of a nonlocal interaction, we have to make the effective
chiral action gauge invariant.  In fact, this gauge-invariant problem
for the nonlocal chiral quark model from the instanton vacuum 
has been already treated in Refs.~\cite{Musakhanov:2002xa,
Kim:2004hd,Nam:2007gf} to which we refer for details.  When the
external gauge field is weak, we can simply replace the usual
derivatives by the covariant ones for soft external vector and
axial-vector fields.  Thus, note that the modified derivative inside
$\sqrt{M}$ has been also replaced by
$i\rlap{/}{\bar{\partial}}+\gamma_5\gamma^{\mu}\frac{\tau^a}{2}J^a_{5\mu}$. 

Using the Lehmann-Symanzik-Zimmermann (LSZ) reduction formula
and Eq.~(\ref{eq:PCAC}), one can obtain the following
expression~\cite{Christov:1995vm}:  
%EQUATION>>>
\begin{eqnarray}
\label{eq:LSZ}
i\sqrt{2}\delta^{ab}F_\pi(q^2,\mu)q_{\mu}
&=&\mathcal{K}_\pi\int d^4x \langle 0| T\,[A^a_{\mu}(x)
\pi^b_\mathrm{phy} (0)]| 0 \rangle\,e^{iq\cdot x} 
\cr
&=&\frac{\mathcal{K}_\pi}{C_r(\mu)}
\int d^4x\langle0|T\,[A^a_{\mu}(x)\pi^b(0)]|0\rangle\,e^{iq\cdot x},
\end{eqnarray}
%EQUATION<<<
where $\mathcal{K}_\pi$ denotes the inverse of the pion propagator,
$\mathcal{K}_\pi=q^2+m^{*2}_\pi$,  in which the asterisk designates
the density modification.  The physical pion weak decay constant is 
defined at $q^2\to-m^{*2}_\pi$.  Eq.~(\ref{eq:LSZ}) can be further
evaluated by the second functional derivative of 
Eq.~(\ref{eq:ECA}) with respect to the $J^a_{5\mu}$ and 
source field $J^b_5$ for the pion: 
%EQUATION>>>
\begin{equation}
\label{eq:LSZ1}
\langle0|T[A^a_{\mu}(x)\pi^b(0)]|0\rangle=
\frac{\delta^2\ln\mathcal{Z}_\mathrm{eff}[\pi,\mu,J^a_{5\mu}]}
{\delta J^a_{5\mu}(x)\,\delta J^b_5(0)}=\int d^4z
\frac{\delta^2\mathcal{S}_\mathrm{eff}[\pi,\mu,J^a_{5\mu}]}
{\delta J^a_{5\mu}(x)\,\delta \pi^b(z)}\mathcal{K}^{-1}_\pi(z).
\end{equation}
%EQUATION<<<
Having performed some tedious calculation, we arrive at  
%EQUATION>>>
\begin{eqnarray}
\label{eq:Fpi}
F_\pi(\mu)P_{\mu}&=&\frac{4N_c}{C_rF_0}\int\frac{d^4k}{(2\pi)^4}
\Bigg[\underbrace{
\frac{\sqrt{M(\bar{k})M(\bar{k}-P)}
[(P_{\mu}-\bar{k}_{\mu})M(\bar{k})+\bar{k}_{\mu}M(\bar{k}-P)]}
{[\bar{k}^2+M^2(\bar{k})][(\bar{k}-P)^2
+M^2(\bar{k}-P)]}}_\mathrm{local\,\,cont.}
\cr
&-&\underbrace{\frac{M(\bar{k})\sqrt{M(\bar{k})}\sqrt{M(\bar{k}-P)}_{\mu}
-M(\bar{k})\sqrt{M(\bar{k})}_{\mu}\sqrt{M(\bar{k}-P)}}
{\bar{k}^2+M^2(\bar{k})}}_\mathrm{nonlocal\,\,cont.}\Bigg],
\end{eqnarray}
%EQUATION<<<
where $\bar{k}=(\vec{k},k_4+i\mu)$ and $C_r=C_r(\mu)$.  We also have
used the following expression: 
%EQUATION>>>
\begin{equation}
\label{eq:deri}
\sqrt{M(k)}_{\mu}=\frac{\partial\sqrt{M(k)}}{\partial k_{\mu}}.
\end{equation}
%EQUATION<<<
Note that $F_\pi$ in Eq.~(\ref{eq:Fpi}) consists of two contributions,
that is, the local (L) and nonlocal (NL) ones.  The later contains
derivatives of $M(\bar{k})$, while the first not.  Using
Eq.~(\ref{eq:Fpi}) and relevant expansions with respect to the
momentum given in Appendix, we can compute $F_\pi^s$ and $F_\pi^t$
separately.  The local contributions are obtained as 
%EQUATION>>>
\begin{eqnarray}
\label{eq:FLs}
F^s_{\pi,\mathrm{L}}(\mu)
&=&\frac{4N_c}{C_rF_0}\int\frac{d^4k}{(2\pi)^4}
\frac{1}{(k^2+\mathcal{M}^2)^2}
\left[\mathcal{M}^2-\frac{1}{2}k^2\mathcal{M}\tilde{\mathcal{M}}'
-5\mu^2\,k^2_4\tilde{\mathcal{M}}'^2\right],
\\
\label{eq:FLt}
F^t_{\pi,\mathrm{L}}(\mu)
&=&
\frac{4N_c}{C_rF_0}\int\frac{d^4k}{(2\pi)^4}
\frac{1}{(k^2 + \mathcal{M}^2)^2}
\left[\mathcal{M}^2-\frac{1}{2}k^2\mathcal{M}\tilde{\mathcal{M}}'
-\mu^2\,k^2_4\tilde{\mathcal{M}}'^2\right],
\end{eqnarray}
%EQUATION<<<
where $\mathcal{M}=\mathcal{M}(k)$ is the momentum-dependent quark
mass in free space (see Appendix). In deriving Eqs.~(\ref{eq:FLs}) and
(\ref{eq:FLt}), we have assumed the soft pion, {\it i.e.} 
$P_4\ll\frac{1}{\bar{\rho}}\approx600$ MeV and $P^2=m^2_\pi= 0$.
From Eq.~(\ref{eq:FLt}), one can easily see that the time component must
be larger than the space one for the local contribution.  Similarly,
the nonlocal contributions can be evaluated as follows: 
%EQUATION>>>
\begin{eqnarray}
\label{eq:FNLs}
F^s_{\pi,\mathrm{NL}}(\mu)
&=&-\frac{4N_c}{C_rF_0}\int\frac{d^4k}{(2\pi)^4}
\frac{1}{k^2+\mathcal{M}^2}
\left[\mathcal{M}\tilde{\mathcal{M}}'
+\frac{1}{2}k^2 \mathcal{M}\tilde{\mathcal{M}}''
-\frac{1}{2}k^2\tilde{\mathcal{M}}'^2
-4\mu^2k^2_4\tilde{\mathcal{M}}'\tilde{\mathcal{M}}''
\right],
\\
\label{eq:FNLt}
F^t_{\pi,\mathrm{NL}}(\mu)
&=&F^s_{\pi,\mathrm{NL}}(\mu).
\end{eqnarray}
%EQUATION<<<
While the time and space components are different each other for the
local contribution, they turn out to be the same for the nonlocal one
at the leading order because of the soft pion. 

When $\mu$ is switched off, the time component equals the space
one, {\it i.e.} $F^s_\pi = F^t_\pi$ as expected, and the analytic expression for
$F_\pi(0)$ leads to the following: 
%EQUATION>>>
\begin{equation}
\label{eq:Fpifree}
F_\pi(0)=\frac{4N_c}{C_rF_0}\int\frac{d^4k}{(2\pi)^4}\left[
\frac{\mathcal{M}^2-\frac{1}{2}k^2\mathcal{M}\tilde{\mathcal{M}}'}
{(k^2+\mathcal{M}^2)^2}
-\frac{\mathcal{M}\tilde{\mathcal{M}}'
+\frac{1}{2}k^2 \mathcal{M}\tilde{\mathcal{M}}''
-\frac{1}{2}k^2\tilde{\mathcal{M}}'^2}{k^2+\mathcal{M}^2}\right],
\end{equation}
%EQUATION<<<
which is already obtained in several works, for example, in
Refs.~\cite{Bowler:1994ir,Franz:1999ik}. The $C_rF_0$ can be
determined within the present framework, resulting in
$C_rF_0\approx{F^\mathrm{exp}_\pi}\approx93$ 
MeV~\cite{Nam:2006ng}. 

The quark-number chemical potential $\mu$ being turned on, the
field-renormalization constant $C_r$ may in principle have some
modifications due to medium effects.  Hence, we modify
Eq.~(\ref{eq:Fpi}) simply, replacing it by 
$C_r(\mu)F_0\to F_\pi(\mu)$ in the denominator.  Finally, we arrive at 
the following expressions for the $F^s_\pi$ and $F^t_\pi$ from
Eq.~(\ref{eq:Fpi}):  
%EQUATION>>>
\begin{equation}
\label{eq:Fpi1}
F^2_\pi(\mu)P_{\mu}
=\left([F^s_\pi(\mu)]^2P_i,[F^t_\pi(\mu)]^2P_4\right),
\end{equation}
%EQUATION<<<
where
%EQUATION>>>
\begin{eqnarray}
\label{eq:}
[F^s_\pi(\mu)]^2&=&[F^\mathrm{exp}_\pi]^2+4N_c\,\mu^2
\int\frac{d^4k}{(2\pi)^4}\left( 
\frac{4k^2_4\tilde{\mathcal{M}}'\tilde{\mathcal{M}}''}{k^2+\mathcal{M}^2}
-\frac{5k^2_4\tilde{\mathcal{M}}'^2}{[k^2+\mathcal{M}^2]^2}
\right),
\cr
[F^t_\pi(\mu)]^2&=&[F^s_\pi(\mu)]^2+4N_c\,\mu^2
\int\frac{d^4k}{(2\pi)^4}
\frac{4k^2_4\tilde{\mathcal{M}}'^2}{[k^2+\mathcal{M}^2]^2},
\end{eqnarray}
%EQUATION<<<
which we will solve numerically in the next Section.
%-------------------------------------------------
\section{Numerical results}
\begin{figure}[t]
\begin{tabular}{cc}
\includegraphics[width=7.5cm]{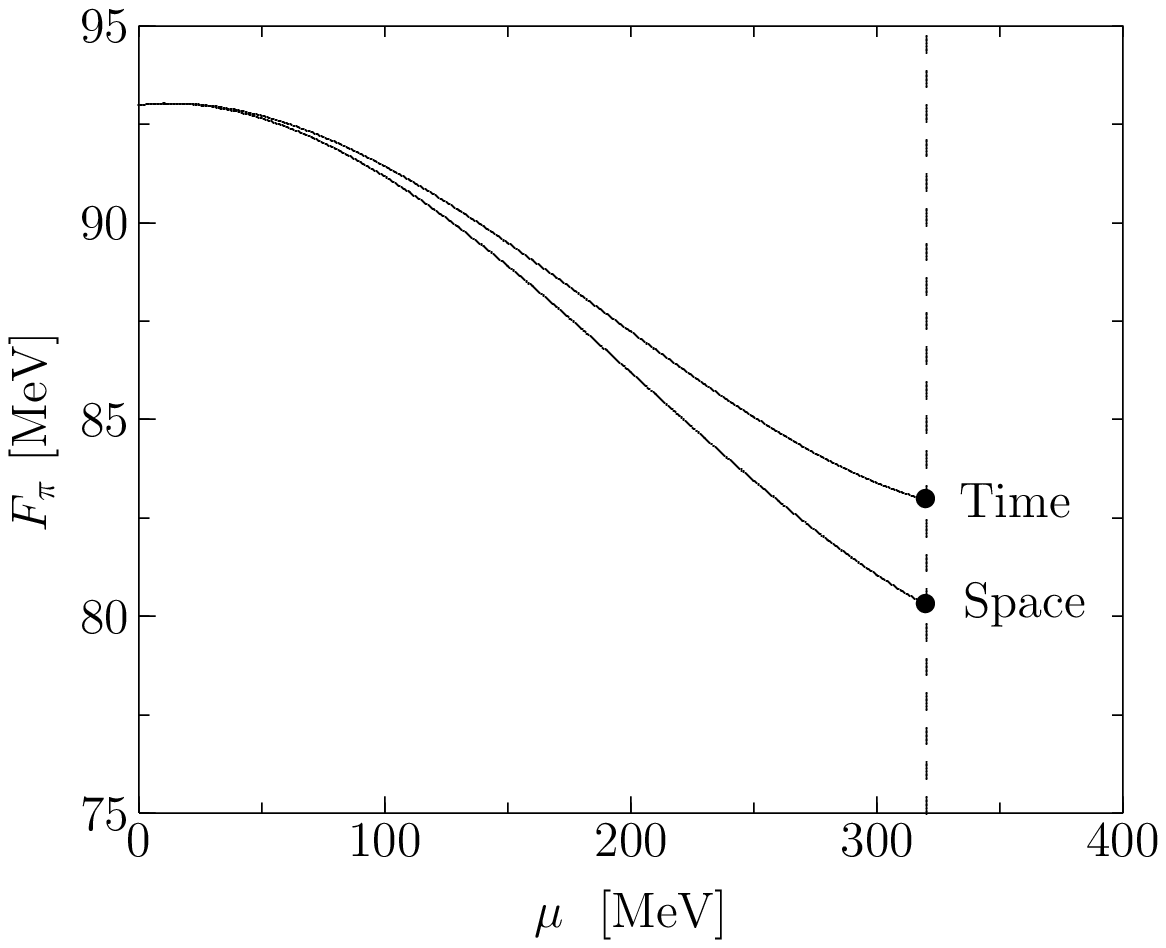}
\includegraphics[width=7.5cm]{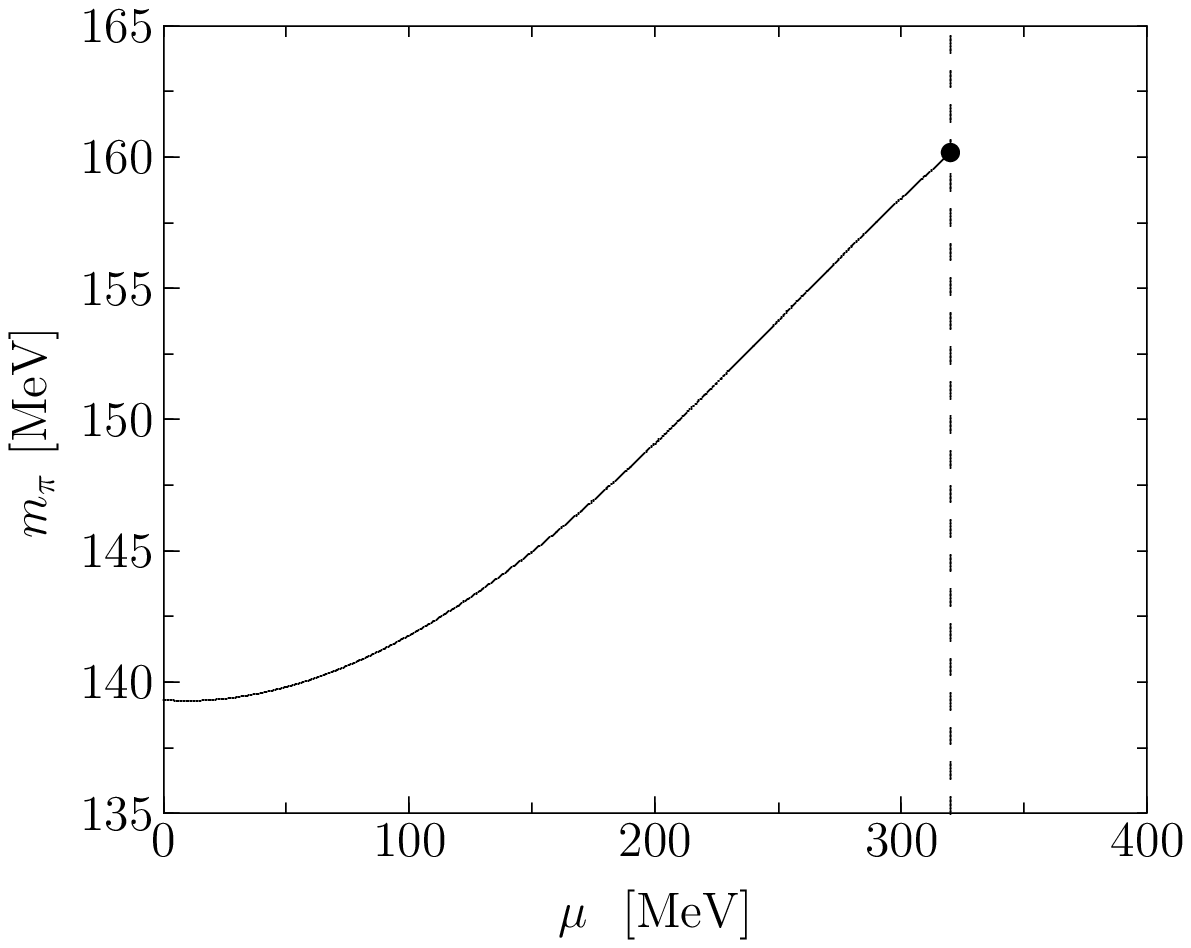}
\end{tabular}
\caption{$F^s_\pi$ and $F^t_\pi$ (left), and $m_\pi$ (right) as
  functions of the quark-number chemical potential $\mu$ up to
  $\mu=\mu_c\approx320$ MeV.}        
\label{fig1}
\end{figure}

%-------------------------------------------------
We now present numerical results for the pion weak decay
constant at finite density.  In the left panel of Fig.~\ref{fig1}, we
show $F^s_\pi$ and $F^t_\pi$ as functions of $\mu$. As shown in
Fig.~\ref{fig1}, the time component of $F_\pi$ is larger than that of
the space one, whereas both of them decrease smoothly with respect to
$\mu$.  At $\mu=0$ we find $F^s_\pi=F^t_\pi\approx93$ MeV as it should
be, and we obtain $F^t_\pi\approx 82.96$ MeV and $F^s_\pi\approx
80.29$ MeV at $\mu=\mu_c\approx 320$ MeV.  When we examine the ratio
$F^{(s,t)}/F^\mathrm{exp}_\pi$, it must be unity at $\mu=0$ and then
it is getting smaller gradually as $\mu$ increases.  At the critical
density ($\mu_c$), it turns out that $F^t_\pi/F^\mathrm{exp}_\pi
\approx 0.89$ and $F^s_\pi / F^\mathrm{exp}_\pi \approx 0.86$. From
these observations, $F^s_\pi/F^t_\pi$ is less than unity
for the whole region of the NG phase, and the $F_\pi$ is reduced by
about $13\sim16\%$.  We summarize the results in Table~\ref{table1}.  

We are now in a position to discuss our results in comparison to those
from other theoretical approaches.  In the QCD sum
rule~\cite{Kim:2003tp}, it was discussed that the splitting between 
the time and space components is represented by the dimension-five
condensates at finite density.  However, the computed ratios are
smaller than ours by about $10\,(20)\%$:
$F^t_\pi/F^\mathrm{exp}_\pi\approx0.79\,(0.69)$ and
$F^s_\pi/F^\mathrm{exp}_\pi\approx0.78\,(0.68)$ for $m_\pi=139\,(159)$
MeV at normal nuclear matter density $\rho_0=0.17\,\mathrm{fm}^{-3}$
corresponding approximately to $\mu \approx 300$
MeV~\cite{Buballa:2003qv}.  Especially, $F^s_\pi/F^\mathrm{exp}_\pi$
becomes much smaller, when the intermediate $\Delta$ state is
considered ($0.78\to0.57$).  The ratios were also studied in in-medium
$\chi$PT in the heavy baryon
limit~\cite{Kirchbach:1997rk,Meissner:2001gz}: It was 
found that $F^t_\pi/F^\mathrm{exp}_\pi\approx0.90$, which is
compatible with the present results, whereas the space component turns
out to be much smaller: $F^s_\pi/F^\mathrm{exp}_\pi\approx0.25$ at
$\rho_0$.   

We would like to discuss more on the splitting between $F^s_\pi$ and
$F^t_\pi$, since we have not observed such a large difference between
$F^s_\pi$ and $F^t_\pi$ in the present framework.  For instance, in
Ref.~\cite{Kirchbach:1997rk}, the analytic expressions for $F_\pi$ at
finite density are given from the axial-vector and pseudoscalar
correlators, being developed at the tree level as follows: 
%EQUATION>>>
\begin{eqnarray}
\label{eq:kirch}
F^s_\pi(\rho_0)&=&\left[1+\frac{2c_3\rho_0}{(F^\mathrm{exp}_\pi)^2}
\right] \left[1-\frac{\Sigma_{\pi N}\,\rho_0}{(F^\mathrm{exp}_\pi)^2
    m^2_\pi}\right]^{-1}, \cr
F^t_\pi(\rho_0)&=&\left[1 + \frac{2(c_2+c_3)
    \rho_0}{(F^\mathrm{exp}_\pi)^2} \right]
\left[1-\frac{\Sigma_{\pi N}\,\rho_0}{(F^\mathrm{exp}_\pi)^2
    m^2_\pi}\right]^{-1}, 
\end{eqnarray}
%EQUATION<<<
where $c_2$ and $c_3$ stand for the coefficients of the effective
chiral pion-nucleon  Lagrangian in the heavy-baryon limit,
corresponding to the terms $(v\cdot\partial\pi)^2$ and
$(\partial\pi)^2$, respectively. Here, $v_\mu$ is the four velocity of the
heavy baryon, whereas $\Sigma_{\pi N}$ denotes the nucleon
$\Sigma$-term. The values of $c_2$ and $c_3$ are estimated for
instance from the low-energy pion-nucleon scattering data,
isospin-even scattering length. Note that these coefficients contain
information on the $p$-wave contribution such as that of the $\Delta$ 
state~\cite{Bernard:1996gq}.  Generally, the value of the ratio
$(c_2+c_3)/c_3$ is about $\frac{1}{2}$ with sign difference: $c_3<0$
and $c_2>0$~\cite{Kirchbach:1997rk,Park:2001ht}.  Because of this
sizable difference between $c_2$ and $c_3$, one observes $F^s_\pi\ll 
F^t_\pi$ as similar to that of the QCD sum rule calculation with the
intermediate $\Delta$-state contribution.  Thus, this discrepancy from
the present work may be due to the fact that the present work is only
based on the quark-pion degrees of freedom.

Reference~\cite{Weise:2001sg}, in which the GOR relation was used for
the time component of the axial-vector current at finite density,
gives the analytic expression for the ratio as 
%EQUATION>>>
\begin{equation}
\label{eq:WEI}
\frac{F^t_\pi(\rho_0)}{F^\mathrm{exp}_\pi}\approx\left[
1-\frac{\Sigma_{\pi N}\,\rho_0}{(F^\mathrm{exp}_\pi)^2m^2_\pi}
\right]^{\frac{1}{2}},
\end{equation}
%EQUATION<<<
where $\rho$ and $\Sigma_{\pi N}$ are the nuclear matter density and
$\pi N$ $\Sigma$-term chosen to be $50$ MeV, respectively.  The
corresponding result is about $0.82$ at $\rho=\rho_0$.  

%TABLE>>
\begin{table}[b]
\begin{tabular}{c|c|c|c}
&$F^s_\pi$&$F^t_\pi$&$m_\pi$\\
\hline
$\mu=0$&$93$ MeV&$93$ MeV&$139.33$ MeV\\
\hline
$\mu=\mu_c\approx320$ MeV&$80.29$ MeV&$82.96$ MeV&$160.14$ MeV\\
\hline
Modification&$16\%\downarrow$&$13\%\downarrow$&$15\%\uparrow$\\
\end{tabular}
\caption{$F^s_\pi$, $F^t_\pi$ and $m_\pi$ at finite density.}
\label{table1}
\end{table}
%TABLE<<<

Now, we would like to investigate the change of the pion mass at
finite density.  Note that in the present work we do not distinguish
the pion by its charge, so that the result is charge-independent.  The 
$m_\pi$ must vanish in the chiral limit, which is the case of the
present work.  However, as shown in  
Refs.~\cite{Diakonov:1985eg,Musakhanov:1998wp,Musakhanov:2001pc,
Nam:2006ng}, the light current-quark mass ($m_q\lesssim 5$ MeV) can
easily be included within the present framework.  Hence, we now
extend the results for $F^{s,t}_\pi$ slightly beyond the chiral limit.
For this purpose, we utilize the in-medium GOR relation~\cite{Thorsson:1995rj,Wirzba:1995sh}:  
%EQUATION>>>
\begin{equation}
\label{eq:GOR}
(m^*_\pi F^t_\pi)^2=2m_q\langle{iq^{\dagger}q}\rangle^*,
\end{equation}
%EQUATION<<<
where $m_q$ is the current quark mass, taken to be around $5$
MeV.  As for the chiral condensate, we use Eq.~(\ref{eq:CC}) as
obtained previously~\cite{Nam:2008ff}.  In the right panel of
Fig.~\ref{fig1}, we draw $m_\pi$ as a function of $\mu$. In free
space, we observe $m_\pi=139.33$ MeV which is in good agreement 
with the experimental value, whereas $m_\pi=160.14$ MeV at the
critical value $\mu\approx 320$ MeV.  This observation tells us that
the pion mass increases almost linearly and  at $\mu_c$ it becomes
about $15\%$ heavier than that in free space.  We summarize this
result in Table~\ref{table1}.  The pion mass at finite density was
also investigated within the QCD sum rule~\cite{Kim:2003tp},
meson-baryon $\chi$PT~\cite{Kaiser:2001bx,Delorme:1996cc,Park:2001ht}
and $s$-wave pion-nucleus phenomenological potential 
models~\cite{Waas:1997kg,Friedman:1998ed}.  In these works,
it turned out that the pion mass increases by $5\sim20\%$ at 
normal nuclear matter density which is qualitatively consistent with
our result, {\it i.e.} it increases by about $15\%$.  However,
Ref.~\cite{Meissner:2001gz} has shown that the mass of the positive
charged pion decreases as a function of nuclear density in finite nuclei or inÊasymmetric (neutron-dominated) nuclear matter, while the mass of the neutral pion does hardly change at all.  
%FIGURE>>>
%FIGURE<<<

%-------------------------------------------------
\section{Summary and conclusion}
%-------------------------------------------------
We have investigated the pion weak decay constant and pion mass at 
finite density within the framework of the nonlocal chiral quark model 
from the instanton vacuum in the presence of the finite quark-number 
chemical potential.  The critical value of $\mu$ was determined in the
present framework consistently.  The Nambu-Goldstone phase survives
till $\mu=\mu_c\approx 320$ MeV, then the first-order phase transition
takes place into the color-superconducting phase.  The
medium-modified effective chiral action being used, the pion weak
decay constant was computed from the pion-to-vacuum transition matrix
element.  Due to the breakdown of Lorentz invariance at finite
density, the time and space components of the pion weak decay constant
were obtained separately.  In the calculation, we assumed the soft
pion.  

As the final results, we obtained $F^t_\pi=82.96$ MeV and
$F^s_\pi=80.29$ MeV at the critical value of $\mu$.  These values were
in qualitative agreement with those of other theoretical models.
Considering the in-medium Gell-Mann-Oakes-Renner relation, we also
studied the pion mass modification at finite density, employing the
present result of the pion weak decay constant and the previously
calculated chiral condensate.  We found that the pion mass $m_{\pi^+}$
increased almost linearly with respect to $\mu$ and at the critical
value of $\mu$ it becomes about $15\%$ larger than the free-space
value: $139.33\to160.14$ MeV.  

In general, we conclude within the present framework that the order of
the medium modification for $F_\pi$ and $m_\pi$ is altogether about 
$10\sim20\%$ at the critical value of $\mu$.   We note 
that this consequence is not much different from other model
estimations.  However, the splitting between the time and space
components of $F_\pi$ has turned out to be relatively small in
comparison to those in in-medium chiral perturbation theory and the 
QCD sum rule.  As discussed in the previous sections, this difference
is due to the effects of the intermediate $\Delta$-state contribution
that was not considered in this work.  However, since the present
results are obtained in the strict large $N_c$ limit, we expect that
the $1/N_c$ meon-loop corrections in this framework may contribute to
this splitting.  The corresponding investigation is under way.

%-------------------------------------------------
\section*{Acknowledgment}
%-------------------------------------------------
The authors are grateful to T.~Kunihiro, S.~H. Lee, M.~M.~Musakhanov,
and Y.~Kwon for fruitful discussions.  S.i.N. would like to thank
D.~Jido for discussions especially on the deeply bound pionic atoms.
The present work was supported by Inha University Research Grant
(INHA-37453).  S.i.N. is also grateful to the hospitality of the
Nuclear Theory Group at Inha University during his visit and
acknowledges the partial support from the Inha University Research
Grant  (INHA-37453).  The work of S.i.N. is partially supported by the
Grant for Scientific Research (Priority Area No. 17070002 and
No. 20028005) from the Ministry of Education, Culture, Science and
Technology (MEXT) of Japan.  This work was done under the Yukawa
International Program for Quark-Hadron Sciences. The numerical
calculations were carried out on YISUN at YITP in Kyoto University. 
%-------------------------------------------------
\section*{Appendix}
%-------------------------------------------------
The momentum-dependent and density-dependent quark mass can be
expanded as follows:
%EQUATION>>>
\begin{eqnarray}
\label{eq:1}
&&M(\bar{k})\approx\mathcal{M}(k)+2i\mu k_4\tilde{\mathcal{M}}'(k),
\,\,\,\,M(\bar{k}-P)\approx\mathcal{M}(k)+
2(i\mu k_4-k\cdot P)\tilde{\mathcal{M}}'(k),
\cr
&&\sqrt{M(\bar{k})}\approx\sqrt{\mathcal{M}(k)}
+i\mu k_4\frac{\tilde{\mathcal{M}}'(k)}{\sqrt{\mathcal{M}(k)}},
\cr
&&\sqrt{M(\bar{k}-P)}\approx\sqrt{\mathcal{M}(k)}
+(i\mu k_4-k\cdot P)\,\frac{\tilde{\mathcal{M}}'(k)}{\sqrt{\mathcal{M}(k)}},
\cr
&&\frac{1}{\sqrt{M(\bar{k})}}\approx\frac{1}{\sqrt{\mathcal{M}(k)}}\left[1
-i\mu k_4\frac{\tilde{\mathcal{M}}'(k)}{\mathcal{M}(k)}\right],
\cr
&&\frac{1}{\sqrt{M(\bar{k}-P)}}\approx\frac{1}{\sqrt{\mathcal{M}(k)}}
\left[1-(i\mu k_4-k\cdot P)\,\frac{\tilde{\mathcal{M}}'(k)}{\mathcal{M}(k)}
\right],
\cr
&&\frac{\partial M(\bar{k})}{\partial k^2}\approx
\tilde{\mathcal{M}}'(k)+2i\mu k_4\tilde{\mathcal{M}}''(k),
\,\,\,\,\frac{\partial M(\bar{k}-P)}{\partial k^2}\approx
\tilde{\mathcal{M}}'(k)+2(i\mu k_4-k\cdot P)\,\tilde{\mathcal{M}}''(k),
\cr
&&\frac{1}{[\bar{k}^2+M^2(\bar{k})][(\bar{k}-P)^2+M^2(\bar{k})]}
\approx\frac{1}{[k^2+\mathcal{M}(k)]^2}. 
\nonumber
\end{eqnarray}
%EQUATION<<<

The momentum-dependent quark mass is parameterized given below.  Its
derivatives with respect to the momentum are also given as follows:
%EQUATION>>>
\begin{eqnarray}
\label{eq:MD}
\mathcal{M}(k)&=&\frac{4M_0\Lambda^4}{(k^2+2\Lambda^2)^2},
\cr
\tilde{\mathcal{M}}'(k)&=&\frac{\partial \mathcal{M}(k)}{\partial k^2}
=-\frac{8M_0\Lambda^4}{(k^2+2\Lambda^2)^3},
\cr
\tilde{\mathcal{M}}''(k)&=&\frac{\partial \mathcal{M}'(k)}{\partial k^2}
=\frac{24M_0\Lambda^4}{(k^2+2\Lambda^2)^4}.
\nonumber
\end{eqnarray}
%EQUATION<<<
%--------------------------------------------------

\end{document}